\documentstyle[preprint,revtex,12pt]{aps}
\draft
\begin{document}
\bibliographystyle{unsrt}

\begin{title}
 Non-Fermi Liquid Behavior In  Quantum Critical Systems
\end{title}
\author{ Junwu Gan and  Eugene Wong }
\begin{instit}
Department of Physics, The University of British Columbia,	\\
6224 Agricultural Road, Vancouver, B.C. Canada V6T 1Z1
\end{instit}
\date{\today\, \, \small UBCTP-93-007}

\begin{abstract}
The problem of an electron gas interacting via
exchanging transverse gauge bosons is studied
using the renormalization group method. The long wavelength behavior
of  the gauge field
 is shown to be in the Gaussian universality  class
 with a dynamical exponent $z=3$ in dimensions $D \geq 2$.
 This implies that
 the gauge coupling constant is  exactly marginal.
Scattering  of the electrons
by the gauge mode leads to  non-Fermi
liquid behavior in $D \leq 3$.
The asymptotic electron and gauge Green's functions, interaction
vertex, specific heat and resistivity are presented.
%

\end{abstract}



The observation of
 unconventional  normal state properties in the  high Tc
cuprates has
stimulated great interest in $2D$ models possessing a
low energy non-Fermi liquid(NFL) fixed point
and a Fermi surface \cite{ande92,varm89}.  It has been
 realized that NFL behavior would  follow naturally
if the electrons or quasiparticles experience
long range or singular  interactions.
Unfortunately, long range interactions generally do not survive
the screening in the presence of
a large Fermi sea and the low energy
 physics is again a Fermi liquid,
unless they arise from the critical fluctuating mode at a phase
transition where the mass of the mode is tuned to zero.

An exception is
the system of an electron gas interacting
 via exchanging transverse
gauge bosons, like photons \cite{holstein73}.
The interaction cannot be screened because  gauge invariance
prevents the photon from acquiring a mass provided
 gauge invariance is not spontaneously broken.
However, if the gauge field is the regular electromagnetic
one, the effects due to its coupling to the electrons
are suppressed by  the fine structure
constant(1/137) and
the ratio of the  Fermi velocity and the speed of light $v_{F}/c$,
thus practically unobservable.
Recently, the same problem
appeared again in the study
of the half filled Landau level \cite{halperin93} and
 in the context of strongly correlated
systems \cite{lee89,nagaosa90,ioffe89,schofield93}.
The local correlation
 such as eliminating double occupation
induces strong phase fluctuations which may be
described by gauge fields in the long wavelength limit.
 In this case,
the effects of the gauge interaction
 are usually not suppressed. In fact,
it has been suggested that
the gauge interaction is probably an essential
element of an effective theory
of high Tc superconductivity
\cite{lee89,nagaosa90,ioffe89,schofield93}.
Although singular behaviors,
signalling breakdown of the Fermi liquid theory,
have been seen in several physical quantities
for this  system
\cite{holstein73,halperin93,lee89,nagaosa90,ioffe89,schofield93,reizer89,blok93}, and
even some suggestions have been made about
 the  low energy  fixed point \cite{polchinski},
its  nature  still remains  unclear.
In this Letter, we reexamine this
 problem using the renormalization group(RG)
method and derive a scaling
solution of the low energy fixed point.
 In $D \leq 3$, the Fermi liquid characters are
destroyed due to the electron scattering off the gauge mode
leading to
a power divergence in the electron spectral density.

We consider the following Hamiltonian,
\begin{equation}
H=\int d^{D}r  \; \psi^{\dagger}(\vec{r}) \left[
\frac{1}{2 m} (-i \vec{\nabla} - g \vec{A} )^{2}
 - \mu \right] \psi(\vec{r})
+ \frac{1}{2} \int  d^{D}r
\left[ \left( \frac{\partial \vec{A} }{\partial t} \right)^{2}
+  \left(  \vec{\nabla} \times  \vec{A}  \right)^{2}
 \right] ,
\end{equation}
where $\psi$ and
 $ \psi^{\dagger}$ are electron
annihilation and creation operators
with spin index neglected, $\mu$ is the chemical potential, and
$\vec{A}$ is the transverse vector potential in the Coulomb gauge.
 We do not
include the scalar potential since it is going to be screened.
We set the photon velocity $c=1$ and consider
 $v_{F} \sim c$.
The coupling constant  $g$ is considered
to be less than one but not too small so that the
effects of the gauge interaction
become observable
 at a temperature where other effects, such as impurity
scattering, haven't taken over yet.
We are interested in the low energy and long wavelength behavior
of the system. That is, we shall
 scale the frequency $\nu_{n}$ and the momentum $q$
of the gauge field as well as the  frequency $\omega_{n}$
of the electrons to zero.
But the electron momentum $k$ is
 scaled to the Fermi wave vector $k_{F}$.
This problem is similar
to the quantum critical phenomenon(QCP)
  considered by Hertz \cite{hertz76}.
The only difference is that
in QCP one has to adjust a relevant parameter
 to land on
the critical point. While for the
 gauge interaction, the $T=0$ criticality
is guaranteed by the gauge invariance.

To determine the
low energy and long wavelength  behavior of the gauge
field,  we integrate out the electrons
and  expand the result
in the powers of gauge field $\vec{A}$,
\begin{eqnarray}
S_{\rm eff}(A)&=& S^{(2)}_{\rm eff}(A)
+ \sum \Gamma^{(3)} A^{3}	\nonumber \\
 &+& \sum_{\bar{q}, \bar{q}_{1}, \bar{q}_{2}}
\Gamma^{(4)}_{\alpha\beta\lambda\gamma}
(\bar{q}, \bar{q}_{1}, \bar{q}_{2})
 A_{\alpha}(\bar{q})  A_{\beta}(\bar{q}_{1})
 A_{\lambda}(\bar{q}_{2})
  A_{\gamma}(-\bar{q}-\bar{q}_{1}-\bar{q}_{2})
+ \cdots  ,
\end{eqnarray}
where we have introduced a short hand notation,
 $\bar{q}=(\vec{q},\nu_{n})$.
The $A^{3}$ term has been studied before \cite{fukuyama}.
The quadratic part of the effective action is
\begin{eqnarray}
S^{(2)}_{\rm eff}(A) =
\sum_{\vec{q},\nu_{n}}
\left( q^{2} +  \frac{ \gamma |\nu_{n}| }{q} \right)
\left( \delta_{\alpha\beta}
- \frac{q_{\alpha} q_{\beta} }{ q^{2} } \right)
 A_{\alpha}(\bar{q})  A_{\beta}(-\bar{q}) ,
	\label{S2eff}
\end{eqnarray}
where $ \gamma \sim g^{2} k_{F}^{2} $.
We have dropped the $\nu_{n}^{2}$ term
 in the original action since it is irrelevant under
 the following  scaling
which preserves the form of (\ref{S2eff}),
\begin{equation}
q \rightarrow s q, \hspace{.3in}
 \nu_{n}  \rightarrow s^{3} \nu_{n} ,
\hspace{.2in} {\rm for} \hspace{.1in}  s \rightarrow 0 .
	\label{scaling}
\end{equation}
  Obviously,
the dynamical exponent is  $z=3$.
To see the effects of interactions, we simply
 count scaling dimensions. Under the
 scaling (\ref{scaling}), the gauge field scales
as $ A \sim s^{-(5+D)/2} $.
The $A^{3}$ interaction is marginal if $\Gamma^{(3)}$
scales as $s^{(3-D)/2}$.
{}From Hall effect study \cite{fukuyama}, it has been known
that $\Gamma^{(3)}$ vanishes faster than $s$
 under the scaling (\ref{scaling}). Thus, it is irrelevant
in dimensions $D \geq 2$(transverse
modes exist only in $D \geq 2$).
Similarly, the $A^{4}$ interaction, $\Gamma^{(4)}$,
is marginal if it scales as $\Gamma^{(4)} \sim s^{1-D} $
and so on.
The $\Gamma^{(4)}$ term includes three diagrams (Fig. 1).
We now verify that
 $\Gamma^{(4)}$ is non-singular as $\bar{q} \rightarrow 0$
and furthermore that the constant
term vanishes as required by gauge invariance.
Letting the external
frequencies go to zero
and  then taking the limit  $\vec{q}_{i} \rightarrow 0$,
the leading term in $D=3$ is
\begin{equation}
\Gamma^{(4)}_{\alpha\beta\lambda\gamma}(0)
= \frac{g^{4}}{ 8 m^{2} }
 \delta_{\alpha\gamma} \delta_{\beta\lambda}
\sum_{\vec{k}} \left[
n_{F}'(\epsilon_{\vec{k}})
 + \frac{2 k^{2}}{3m} n_{F}''(\epsilon_{\vec{k}})
 + \frac{ k^{4}}{15m^{2}} n_{F}'''(\epsilon_{\vec{k}})
\right]  = 0 ,	\label{gam40}
\end{equation}
where $n_{F}(\epsilon)$ is the Fermi-Dirac function and the primes
denote derivatives.  The result~(\ref{gam40})  also holds
in other dimensions.
Thus, we reach the conclusion that $\Gamma^{(3)}$, $\Gamma^{(4)}$
and  all
interactions in the effective
gauge action are irrelevant  \cite{hertz76} because higher
order terms are even more suppressed
 under the scaling  (\ref{scaling}).

An immediate consequence of
the irrelevance of all corrections to
the Gaussian action (\ref{S2eff}) is that we can derive the asymptotic
form of the specific
heat.
Carrying out the integration over the gauge field in
(\ref{S2eff}), the free energy is
\begin{equation}
F(T) = \frac{1}{\beta} \sum_{\vec{q},\nu_{n}}
\ln \left( q^{2} +  \frac{ \gamma |\nu_{n}| }{q} \right)
\sim T^{1+D/3} \int^{q_{c}^{3}/\gamma T}_{0} dx \;
 x^{D/3-1}
\int^{\infty}_{0} \frac{dy}{e^{y}-1} \tan^{-1}
 \left(\frac{y}{x}\right) ,
\end{equation}
where in deriving the last expression,
 we cut off the upper limit for
$q$-integration  at $q_{c}$, of order $k_{F}$,
and the frequency integration at $q_{c}^{3}/\gamma$
which has been sent to infinity due to the convergence
of the integration.
It's easy to see that $F(T) \sim T^{2} \ln T$ in $D=3$
and $F(T) \sim T^{1+D/3}$ in $D < 3$. The corresponding
specific heat is $C \sim T \ln T$ in $D =3$ \cite{holstein73},
 and $C \sim T^{D/3}$
in $D < 3$ respectively.
Further corrections have  higher powers in $T$.
These  results
are consistent with the general scaling analysis
since the scaling form of the specific heat is uniquely
determined by the dynamical exponent
and the dimensionality.

Another consequence is that
 the coupling constant $g$ is exactly marginal,
its beta-function vanishes identically.
This follows from the Ward identity which stipulates that
the vertex renormalization factor
$Z_{1}$, in the
standard QED notation, be equal to
the electron wave function renormalization factor $Z_{2}$
which represents the magnitude of
the Fermi surface discontinuity.
 Thus, the renormalization of the coupling constant
$g$ is solely determined by the gauge
field wave function renormalization
factor $Z_{3}$ which remains equal to one because
all corrections due to the interactions are irrelevant.
One also finds a vanishing
beta-function of $g$ by imposing RG invariance on the specific heat
\cite{gan93}.

We now turn to the behavior of the electrons.
In  calculating the electron self-energy, the photon
propagator is given by (\ref{S2eff}).
We do not need to
include  further photon self-energy corrections
because they are irrelevant. Since the coupling
constant $g$ does not flow, we can use perturbation if $g < 1$.
When analytically continuing to the real frequency,
the electron self-energy is
$\Sigma(\vec{k},i\omega_{n}=\omega+i 0^{+})
 = \Sigma'(\vec{k},\omega)
+ i \Sigma''(\vec{k},\omega) $.  To  the
lowest order, we find
\begin{equation}
 \Sigma''(k_{F}, \omega)
= - \frac{ g^{2} v_{F} }{8 \pi}
 \gamma^{D/3-1} |\omega|^{D/3}  ,
	\label{Sigpp}
\end{equation}
The real part of the self-energy is given
 by the Kramers-Kronig relation,
\begin{equation}
  \Sigma'(k_{F}, \omega)
= \frac{2 \omega}{\pi} \;  -\!\!\!\!\!\!\int^{\Omega}_{0}
d\epsilon  \;
\frac{  \Sigma''(k_{F}, \epsilon) }{\epsilon^{2} -\omega^{2}}
\simeq - \frac{  g^{2} v_{F}}{4 \pi^{2} }
 \; \gamma ^{D/3-1} \omega^{D/3}
 -\!\!\!\!\!\!\int^{\Omega/\omega}_{0}
 \frac{dx}{\pi}
 \frac{x^{D/3}}{x^{2}-1}  ,	\label{Sigrel}
\end{equation}
where $\Omega \sim q_{c}^{3}/\gamma$, is the frequency cutoff.

Let us first concentrate on the $D\!=\!3$ case.
{}From (\ref{Sigrel}),
we have
\begin{equation}
Z_{2}(k_{F},\omega)
=\left[ 1- {\partial \Sigma'(k_{F}, \omega) \over \partial \omega}
\right]^{-1}
\simeq 1 - \frac{ g^{2} v_{F} }{4 \pi^{2} }
\ln \left(\frac{\Omega}{\omega}\right) .  \label{z2}
\end{equation}
The physics of this logarithmic term is
similar to the well known infrared catastrophe \cite{anderson67}.
Because of the critical nature of the gauge field,
the electrons near the Fermi surface are dressed by a cascade of
 damped photons.
Technically, the $Z_{2}$ given by (\ref{z2}) is
reliable for
 $\omega > \Omega e^{- 4 \pi^{2}/ ( g^{2} v_{F} )}$.
 In order to find
$Z_{2}$ for $\omega \rightarrow 0 $, we  use
the standard RG method and first obtain
\begin{equation}
\eta = \frac{d \ln Z_{2} }{ d \ln \omega}
=  \frac{ g^{2} v_{F} }{4 \pi^{2} }.			\label{def-eta}
\end{equation}
Bearing in mind  that the coupling constant $g$
stays unchanged, the leading logarithmic series is summed up
by integrating  (\ref{def-eta})
over the range $[\omega, \Omega]$ to
give
\begin{equation}
Z_{2}(k_{F},\omega) \sim \omega^{\eta},
\hspace{.5in}
 G''(k_{F}, \omega) \sim \frac{1}{\omega^{1-\eta}} .
	\label{sumz2}
\end{equation}
The spectral density $G''$ has a power law divergence, removing
all remnant  characters of the quasiparticle and
destroying the Fermi liquid.

In $D < 3$, the situation is less transparent. From (\ref{Sigrel}),
the leading self-energy is
$\Sigma'(k_{F},\omega) \sim \Sigma''(k_{F},\omega)
\sim \omega^{D/3}$,
with an exponent $D/3 < 1$.
As the energy is lowered,
the electron Green's function is dominated
by the effect of the self-energy.
The crucial question is then
 whether or not
more singular terms will appear as $\omega \rightarrow 0$ in higher
order calculations,
such as $\omega^{[1-n(3-D)/3]}$ in the $n$th order.
For the following reason,
we do not expect this kind of terms.
The exact marginality of the coupling constant $g$
in all dimensions means that there should be {\em no}
infrared divergence.
 This does not contradict the appearance
of the logarithmic divergence
in the electron self-energy at $D=3$ which is
purely due to the infrared catastrophe,
indicating  that each electron at
the Fermi surface is accompanied by an infinite number of soft
photons.
The total energy of these photons
is finite. If higher singular powers were generated
in high order calculations for $D<3$ and we still
tried to interpret them as the infrared catastrophe,
it would imply a divergent total energy of the accompanying photons
which is unphysical.
The reasonable expectation
as  suggested by Polchinski \cite{polchinski}
is that once we have included the
new term $\omega^{D/3}$ in the electron Green's function,
there will neither be infrared divergence
 responsible for the renormalization of $g$ nor
infrared catastrophe
which  occurs only  in $D=3$ but
cancels out in physical quantities.
 This is
 partially supported
in the direct evaluation of the first crossing diagram of
the electron self-energy \cite{ioffe93}.
 By finding an asymptotic solution
of the Dyson equations(this approach is physically sensible because
$g$ is exactly marginal),
we verify this expectation. Specifically,
we shall prove that the full gauge propagator
is given by (\ref{S2eff}), and  the full electron propagator,
the irreducible gauge interaction vertex have the following
asymptotic forms for $D <3$,
\begin{eqnarray}
 & & G(\vec{k}, \omega+i\delta) =
\frac{1}{ \lambda_{1} |\omega|^{D/3} {\rm sgn}\omega -
 \epsilon_{\vec{k}} + i \lambda_{2} |\omega|^{D/3} \;
{\rm sgn}\delta } ,
	\label{Gelect}			\\
 & & \Lambda_{\mu}(\bar{k},\bar{k}+\bar{q}, \bar{q}) =
\Lambda \,  k_{F}^{\mu}  ,		\label{vertex}
\end{eqnarray}
where $\lambda_{1}$, $\lambda_{2}$
 and $\Lambda$ are all constants.
Note that in order to assume the scaling form (\ref{Gelect}),
we need $\Sigma(\vec{k},\omega+i\delta)$
for general $\vec{k}$ and $\omega$ which has been calculated
 in \cite{halperin93}. Since
 $\Sigma(\vec{k},\omega+i\delta)$ depends on $\vec{k}$ only
when $\omega < (k-k_{F})^{3}$, the $\vec{k}$
dependence  of $\Sigma(\vec{k},\omega+i\delta)$
is irrelevant under scaling, justifying (\ref{Gelect}).

The three irreducible objects, (\ref{Gelect}), (\ref{vertex})
and the gauge propagator given by (\ref{S2eff}),
have to
 satisfy the Dyson equations (Fig. 2).
At $T=0$, they are
\begin{eqnarray}
 & & \Sigma''(\vec{k},\omega) = - \frac{g^{2} \Lambda}{m^{2}}
\sum_{\vec{q}}  [ k^{2} -(\vec{k}\cdot \hat{q})^{2}]
\int^{0}_{-\omega} \frac{d \nu}{\pi}
G''(\vec{k}+\vec{q},\omega+\nu) D''(q,\nu)  	,
\hspace{.1in} \omega > 0				 \\
 & & \Pi''(\vec{q},\nu) = \frac{g^{2} \Lambda}{m^{2}} \sum_{\vec{k}}
 [ k^{2} -(\vec{k}\cdot \hat{q})^{2}]
\int^{0}_{-\nu} \frac{d \omega}{\pi}
G''(\vec{k},\omega) G''(\vec{k}+\vec{q},\omega+\nu) ,
\hspace{.2in} \nu > 0		\\
 & & \Lambda = 1 - \frac{g^{2} \Lambda^{3} }{m^{2}}
\sum_{\vec{q}}  [ k_{F}^{2} -(\vec{k}_{F} \cdot \hat{q})^{2}]
\int^{0}_{-\infty} \frac{d \nu}{\pi}
\left\{ 2 G'(\vec{k}_{F}+\vec{q},\nu) G''(\vec{k}_{F}+\vec{q},\nu)
D'(\vec{q},\nu)  \right.	\nonumber	\\
  &  &  \hspace{.3in}
+   \left. \left[  G'(\vec{k}_{F}+\vec{q},\nu)^{2}
 -  G''(\vec{k}_{F}+\vec{q},\nu)^{2} \right]
 D''(\vec{q},\nu) \right\}
	+ \cdots			\label{vert-eq}
\end{eqnarray}
where $\Pi''$ is the imaginary part of the photon self-energy.
Note that the vertex equation (\ref{vert-eq})
 contains a series of infinite
skeleton diagrams and only the expressions for
the first two have been written out.
The integrations can be carried out in $D < 3$
keeping only the
leading powers in frequency and momentum. We find
\begin{eqnarray}
 & & \Sigma''(k_{F},\omega) = -
 \frac{ \Lambda g^{2} v_{F}  }{2 (2\pi)^{D-1} }
  \gamma^{D/3-1} \omega^{D/3}   \;
\int^{\infty}_{0}  dx \;  x^{D/3-1}
\ln\left(1+\frac{1}{x^{2}}\right)   	\label{nca}		\\
 & & \Pi''(\vec{q},\nu) =
\frac{\pi \Lambda  g^{2} k_{F}^{D-1} }{2(2\pi)^{D-1}} \;
\frac{\nu}{q}  			\label{Pipp}		\\
 & & \Lambda = 1+ \frac{ \Lambda^{3} g^{2} v_{F} }{ (2\pi)^{D-1} }
\frac{\gamma^{D/3-1} }{ \lambda_{1} }
\int^{\infty}_{0} d x \frac{1}{ (1+x^{2})x^{D/3} } + \cdots  .
		  \label{vertex-f}
\end{eqnarray}
The important point is that (\ref{vertex-f}) is well behaved and no
higher singular terms are generated
in (\ref{nca}) and (\ref{Pipp}).
Strictly speaking,
in order to prove that (\ref{S2eff}), (\ref{Gelect})
and (\ref{vertex})  are the asymptotic
 solution of the Dyson equations
in $D< 3$,
we have to verify that no infrared divergence will be generated in every
skeleton diagram of (\ref{vert-eq}).
Nevertheless, as explained above,
we do not expect divergence in higher order skeleton diagrams
because $g$ is exactly marginal, although they may
contribute to
determining the constants
$\lambda_{1}$, $\lambda_{2}$  and $\Lambda$.
Thus, we conclude that
(\ref{Gelect}), (\ref{vertex})
and the gauge propagator given by (\ref{S2eff})
are indeed the
asymptotic  low energy solution.

{}From (\ref{Gelect}), we see $ G''(k_{F},\omega) \sim \omega^{-D/3}$.
As $D \rightarrow 3$, the exponent is discontinuous
from the $D=3$ value given by (\ref{sumz2}).
It is  then instructive to study  the dimensional crossover
as we  lower the pertinent energy scale, treating
 $D=3-\varepsilon$ as a continuous parameter and
$|\varepsilon| \ll 1$. To analyze (\ref{Sigrel}),
we define  a small energy scale:
$T_{\varepsilon}=\Omega e^{-3/|\varepsilon|}$.
At $\omega > T_{\varepsilon}$, we find
 $ (\gamma \omega)^{-\varepsilon/3} \simeq 1$
and $\Sigma'(\omega)/\omega
\sim  g^{2} v_{F} \ln(\Omega/\omega)$.
Comparing with (\ref{z2}), we see the same behavior
as in $3D$ at $\omega > T_{\varepsilon}$ for
all $\varepsilon$.

At $\omega < T_{\varepsilon}$, we see from (\ref{Sigrel}) that
 the situation is different for
 $D > 3$ and $D < 3$.
In $D>3$,  $ \ln(\Omega/\omega)$ appearing at $\omega > T_{\varepsilon}$
is now cut off by $3/|\varepsilon|$. The frequency dependence
of  $\Sigma'(\omega)/\omega $ dies as $\omega^{|\varepsilon|/3}$.
The system eventually flows to a Fermi liquid like fixed point
with a quasiparticle scattering rate given by (\ref{Sigpp}).
In $D < 3$,  we have $\Sigma'(\omega)/\omega \sim
(\gamma \omega)^{-\varepsilon/3} /\varepsilon$
from direct evaluation.
The effect of  $(\gamma \omega)^{-\varepsilon/3}$
starts to become important  at
$\omega < T_{\varepsilon}$.
As we have argued, there are no other singular terms.
In the numerical prefactor of $(\gamma \omega)^{-\varepsilon/3}$,
each $ \ln(\Omega/\omega)$  appearing
at $\omega > T_{\varepsilon}$
is  again replaced by $3/\varepsilon$
 and the series in $3/\varepsilon$ can be summed up to give a constant
for finite $\varepsilon$. We  thus
recover the $D< 3$ behavior (\ref{Gelect}).
An illustration of the  dimensional crossover
is sketched in Fig. 3.

Although
 the electron Green's function
is gauge dependent,
physical results derived from it are not.
 As an example, we
calculate the resistivity from the Kubo formula.
Since (\ref{vertex}) has no singularity in $D<3$,
the vertex correction in the resistivity does not
change the temperature dependence. So, we find
$\rho \sim T^{D/3}$.
In $D=3$, the electron wavefunction and the vertex corrections have to
be taken into account.
We shall
present the result in future publication.

The similar occurrence of the electron critical scattering at
the quantum phase transition in the presence of
a Fermi surface probably provides the easiest experimental
realization.
The role of the gauge field is then played by the
soft mode of the critical fluctuations.
In the  pressure driven
itinerant ferromagnetic transition, the
critical mode is the  magnon excitation.
  The power law divergence in the electron spectral density
is directly  related to physical observables
in the Fermi surface measurements.
It is interesting to note that many heavy fermion systems
show low temperature critical behavior,
 markedly different from the Fermi liquid
expectation \cite{seam91,tsvelik}.
The critical scattering has  also been seen
at the heavy fermion metamagnetic transition \cite{haen87}.


\acknowledgments
J.G. thanks A. Finkelstein,  A. Schofield,  A. Tsvelik, J. Wheatley
for helpful discussions,
and N. Andrei, P. Coleman for valuable
comments on the manuscript.
 The authors are  grateful to Ian Affleck
for his helpful comments and constant encouragement.
This work was supported in part
by NSERC and CIAR of Canada.

\figure{ The $4$th order interaction vertex of the effective
gauge action. The  thin solid line represents the
non-interacting electron Green's function.}

\figure{Dyson equations. $\Sigma$, $\Pi$
and $\Lambda_{\mu}$ are the irreducible
electron, photon self-energies and
 the irreducible gauge interaction vertex respectively.
The thick lines are full Green's functions. }

\figure{Illustration of the dimensional crossover. For thermodynamic
quantities, the temperature $T$ corresponds $\omega$.}

\end{document}